\documentclass[times,conference]{IEEEtran} 
\usepackage{latex8}
\usepackage{times}
\usepackage{url,fancyhdr,epsfig,graphicx,color,multirow,amsmath,subfigure,slashbox,setspace,algorithmic,algorithm,amssymb}
\usepackage{color}
\newcommand{\rednote}[1]{\color{red}{ \em #1 }\color{black}}
\newcommand{\bluenote}[1]{\color{blue}{ \em #1 }\color{black}}
\renewcommand{\rednote}[1]{}
\renewcommand{\bluenote}[1]{}

\newcommand{\mymax}{\operatorname{max}}

\newcommand{\sinrThresh}{{\rm T}_{SINR}}
\newcommand{\rxSen}{{\rm T}_{RX}}
\newcommand{\mis}{\rm M}
\newcommand{\optiMis}{\widehat {\rm M}}
\newcommand{\edgePresence}{N}
\newcommand{\edgeUnsafe}{\rm U}
\newcommand{\rtsRating}{{\cal {R}}}
\newcommand{\dataRating}{{\cal {D}}}
\newcommand{\dataRatingImp}{\widehat {\dataRating}}
\newcommand{\probMis}{\mu}
\newcommand{\probSrcWin}{\sigma}
\newcommand{\RTSCorrupt}{{\rm K}}
\newcommand{\CTSCorrupt}{\bar {\rm K}}
\newcommand{\ACKCorrupt}{\ddot {\rm K}}
\newcommand{\RTSCorruptImp}{{\widehat {\RTSCorrupt}}}
\newcommand{\CTSCorruptImp}{{\widehat {\CTSCorrupt}}}
\newcommand{\indirectIferEdges}{\rm I}
\newcommand{\maxCumNoiseBelowSen}{\eta}
\newcommand{\cumNoiseFactor}{\nu}
\newcommand{\maxParallel}{\rm P}
\newcommand{\navSensing}{\beta}

\newenvironment{definition}[1][Definition]{\begin{trivlist}
\item[\hskip \labelsep {\bfseries #1}]}{\end{trivlist}}

\begin{document}
\title{The Effect of Scheduling on Link Capacity in Multi-hop Wireless Networks}
\author{Vinay Kolar and Nael B. Abu-Ghazaleh
\ \\
 \begin{minipage}{80mm}
 \begin{center}
        CS. Department, Binghamton University \\
       Binghamton, NY~~13902--6000 \\
   {\{vinkolar,nael\}@cs.binghamton.edu}
\end{center}
\end{minipage}
}
\maketitle

\begin{abstract}
Existing models of Multi-Hop Wireless Networks (MHWNs) assume that
interference estimators of link quality such as observed busy time
predict the capacity of the links.  We show that these estimators do not capture the
intricate interactions that occur at the scheduling level, which have
a large impact on effective link capacity under contention based MAC
protocols.
%; even in
%the presence of high interference, if the scheduling protocol is
%successful in arbitrating the channel, effective operation results. 
We observe that scheduling problems arise only among those interfering
sources whose concurrent transmissions cannot be prevented by the MAC
protocol's collision management mechanisms; other interfering sources
can arbitrate the medium and co-exist successfully.  Based on this
observation, we propose a methodology for rating links and show that
it achieves high correlation with observed behavior in simulation.  We
then use this rating as part of a branch-and-bound framework based on
a linear programming formulation for traffic engineering in static
MHWNs and show that it achieves considerable improvement in
performance relative to interference based models.
\end{abstract}

\section{Introduction}
Multi-hop wireless networks (MHWNs) play an increasingly important
role at the edge of the Internet.  Mesh networks provide an extremely
cost-effective last mile technology for broadband access,
%~\cite{wireless-philadelphia};
%% ( for example, a mesh network providing broadband access to the city
%% of Philadelphia, covering 135 square miles, at less than half the cost
%% of traditional broadband access is currently being
%% built~\cite{wireless-philadelphia}.)
ad hoc networks have many applications in the military, industry and
everyday life,
%~\cite{ref:meshApps}
and sensor networks hold the promise of revolutionizing sensing across
a broad range of applications and scientific disciplines -- they are
forecast to become bridges between the physical and digital worlds.
This range of applications results in MHWNs with widely different
properties in terms of scales, traffic patterns, radio capabilities
and node capabilities.  Thus, effective networking of MHWNs has
attracted significant research interest.

Gupta and Kumar~\cite{ref:gupta_99_capacity} derived the asymptotic
capacity of MHWNs under the assumptions of idealized propagation,
uniformly distributed sources and destinations, and an optimal routing
and packet transmission schedule.  A key observation in deriving this
limit is that the available bandwidth between a pair of communicating
nodes is influenced not only by the nominal communication bandwidth,
but also by ongoing communication in nearby regions of the network
due to interference.

While the asymptotic limit is useful, the analysis cannot be applied
to evaluate, or traffic engineer specific networks and traffic
patterns.  Recently, several efforts to model MHWNs while
incorporating the effect of interference have been carried out
~\cite{ref:jain_03_interference,ref:kodialam_03_interference,ref:kolar_06_mcf}.
One of the important limitations in these works is that they abstract
away scheduling, for example, by ignoring its effect or assuming the
presence of an omniscient scheduler. Existing works in this area use
an estimate of interference to assess the quality of links 
for use in estimating capacity or determining effective globally
coordinated routing configurations.   % However, we show that
% interference estimates provide only a rough approximation of link
% quality and great variations in qualities can arise for the same level
% of interference.  This variation is due to low-level scheduling
% interactions that arise in contention based scheduling protocols such
% as IEEE 802.11. 
 We present these and other related works in
Section~\ref{sec:related}.
%% While the interference on the channel accounts for the spatial
%% relationship between the transmitting links, scheduling interactions
%% contributes to the temporal relationships. 

The first contribution of the paper is to show (in
Section~\ref{sec:correlation}) that commonly used metrics for
capacity provide only an upper bound on the achievable
capacity of a link.  % This includes metrics such, as interference
% level (at the receiver or at the sender) and average utilization of
% the channel. 
Scheduling effects, which are not accounted for by these metrics, play
an important, often defining, role in determining the effective
capacity of a link.  The effects arise from the inability of practical
contention based MAC protocols, such as IEEE 802.11, to prevent
collisions in some cases.  We show that metrics such as observed
percentage of MAC level timeouts, can estimate the scheduling
effect on capacity, and when combined with the interference metrics
provide accurate estimates of link quality.
% link
% quality and achievable throughput.  This argument is based on two
% observations: (1) collisions lead to lost transmissions and increased
% backoff, which reduce the capacity of the network; and (2) The
% overwhelming majority of collisions (and hence timeouts) occur due to
% unmanaged contention that cannot be prevented by the MAC protocol.
% Many links that interfere with each other can handshake successfully
% using CSMA and CA and do not exhibit a high number of collisions;
% collisions are indicative of destructively interacting links.

It is not readily clear how to estimate the expected scheduling
interactions among a given set of active links.  The second
contribution of the paper is an analysis of this problem and a
proposed model for estimating the scheduling effects.  We build on the
observation that {\em virtually all collisions are caused only by
  the subset of interfering sources whose concurrent transmission
  cannot be prevented by the MAC protocol}, due to the imprecise
nature of collision prevention mechanisms in contention based MAC;
other interfering sources are able to handshake effectively and avoid
collisions.  We develop a model for identifying these sets of nodes
and predicting the quality of a given link.  We show that proposed
metrics effectively predict the expected percentage of MAC timeouts.
% Thus, the IBLR metric can be used in conjunction with interference
% metrics for accurate characterization of link quality.  
The analysis and the development of scheduling aware metrics are
presented in Section~\ref{sec:schedFormulation}.

To demonstrate the effectiveness of the proposed metrics, we extend a
linear programming formulation for global route planning in MHWNs to
use the scheduling-aware metrics.  % Routes are first developed with
% respect to the interference objective function.  Each configuration is
% then evaluated using the scheduling-aware metrics, and links with
% mutually destructive interference are identified.  These interactions
% are then taken into account in the model as additional constraints and
% the routes are re-planned.  
We show that the resulting routes achieve large performance
improvements relative to routes obtained considering interference
metrics only; the improvement is even larger with respect to a greedy
routing protocol such as DSR.  The algorithm is presented in
Section~\ref{sec:routing} and the experimental evaluation is presented
in Section~\ref{sec:schedRes}.
%Discussion of the results is
%presented in Section~\ref{sec:discussion}.  
%Finally,
%Section~\ref{sec:conclude} presents some concluding remarks

% \rednote{We should moderate this next paragraph.  Perhaps state the
%   utility of the model and then discuss extensions to it to account
%   for these other things that we dont have now} \bluenote{Vinay: Does
%   the below para look fine now?}  

% Although realistic wireless propagation environments is not modeled,
% the study provides a deeper understanding about the interaction in
% using a statistically representative wireless propagation model. 
The centralized approach of the model is useful for analysis and for
static networks (e.g., mesh networks), but not directly for dynamic networks.
% % The developed approach uses idealized wireless propagation models.  It
% % is well known that realistic wireless propagation has significant
% % implications on the behavior and performance of wireless networking
% % protocols.  
% % In addition, the approach is centralized; such an approach
% % is useful for analysis and for static networks (e.g., mesh networks),
% % but not for dynamic networks.  
However, we believe that identifying the nature and scope of
interactions provides an excellent starting point for developing
distributed algorithms that result in effective routing configurations
in dynamic networks.  We present our conclusions and future research
directions in Section~\ref{sec:conclude}.

%\rednote{Need related work section here. Starting point can be from the proposal.  This comment is in the scheduling.tex  file}
%\input{background} %2
\section{Related work}
\label{sec:related}
A primary challenge of efficient channel usage for the emerging
applications in MHWNs would be to identify the critical metrics 
of routing protocol that enables to approximate an optimal 
route configuration.Under idealized assumptions, 
hop count has been experimentally shown to 
determine a stand-alone connection performance metric
~\cite{ref:li_01_capacity}. The validity of hop-count as a sole 
metric of path quality is brought into question because it fails to account for two
crucial factors: the channel state between the sender and receiver
(assumed perfect within transmission range) and interference from
other connections (which is ignored when taking routing decisions).
It does not capture the underlying MAC and physical layer which has a great 
effect on the routing protocol~\cite{ref:barrett_03_interactions, ref:takai_01_phy}. 
Using the shortest number of hops often lead to the choice of distant next-hop,
thus reducing signal strength and channel quality between the
hops~\cite{ref:decouto_02_shortestPath}.
Researchers have studied approaches for estimating the
quality of links including Round Trip Time
(RTT)~\cite{ref:adya_04_multiradio} and the expected number of
retransmissions (ETX)~\cite{ref:decouto_03_metric}. Draves et al.
carried out a comparison of these and other 
metrics~\cite{ref:draves_04_comparison} and concluded that ETX is the
most effective measure in their experimental static network. We believe
that the scheduling interactions analyzed in the paper would reason
the superior performance of ETX.

Various approaches and metrics has resulted in studies that not only 
identifies the parameters and their relationships but also quantifies them. 
Most of the studies use the notion of \textit{conflict graphs} and 
\textit{maximal independent sets} for capturing the wireless effects.  
Jain et al. and Kodialam et al.
~\cite{ref:jain_03_interference,ref:kodialam_03_interference} propose an 
interference model and show the effect of interference on the aggregate throughput. 
However, they assume all-or-none interference and an optimal scheduler. 
While such interference aware routes \textit{spatially}
characterize the routes, the scheduling interactions studied in this paper 
capture the \textit{temporal} properties of the routes. 
% Garetto et al.
% ~\cite{ref:garetto_06_starve} model the problem 
% of starvation in multi-hop wireless network. While
% the above study models scheduling using busy time probabilities and estimates
% starvation, the objective of our study is to capture the low level interactions 
due to contention based scheduling.

Link behavior and routing protocols that account for realistic
wireless channel state has been empirically studied
in~\cite{ref:ganesan_02_complex,ref:woo_03_taming,ref:cerpa_05_temporal,
  ref:padhye_05_estimateInterference}.  Garetto et
al~\cite{ref:garetto_06_starve} estimate the effect of scheduling on
the throughput of CSMA channels.  They use a similar approach to ours
in computing maximal cliques to identify concurrently transmitting
nodes.  However, unlike our approach, they use a Markovian model to
estimate the scheduling effects statistically and use the simplified
protocol model.  In contrast, our model is more constructive in nature
starting from the underlying causes of collisions.
\section{Link Capacity Metrics}
\label{sec:correlation}

In this section, we first show using a simple experiment that
aggregate interference metrics provide only a coarse estimate of link
quality, especially under medium to high interference, because they
ignore the effect of scheduling.  We simulate different sets of 144
uniformly distributed nodes with 25 arbitrarily chosen {\em one-hop}
connections.  The aim of this experiment is to show how the
performance achieved by these connections correlates with existing
capacity metrics.  We chose one hop connections to eliminate multi-hop
artifacts such as self-interference and pipelining and isolate
link-level interactions.  Each source sends CBR data at a rate high
enough to ensure that it always has packets to send.

\begin{figure*}
\begin{center}
        \mbox{
         \subfigure[\label{fig:throughSimBusySrcHirate}]{\includegraphics[scale=0.25]{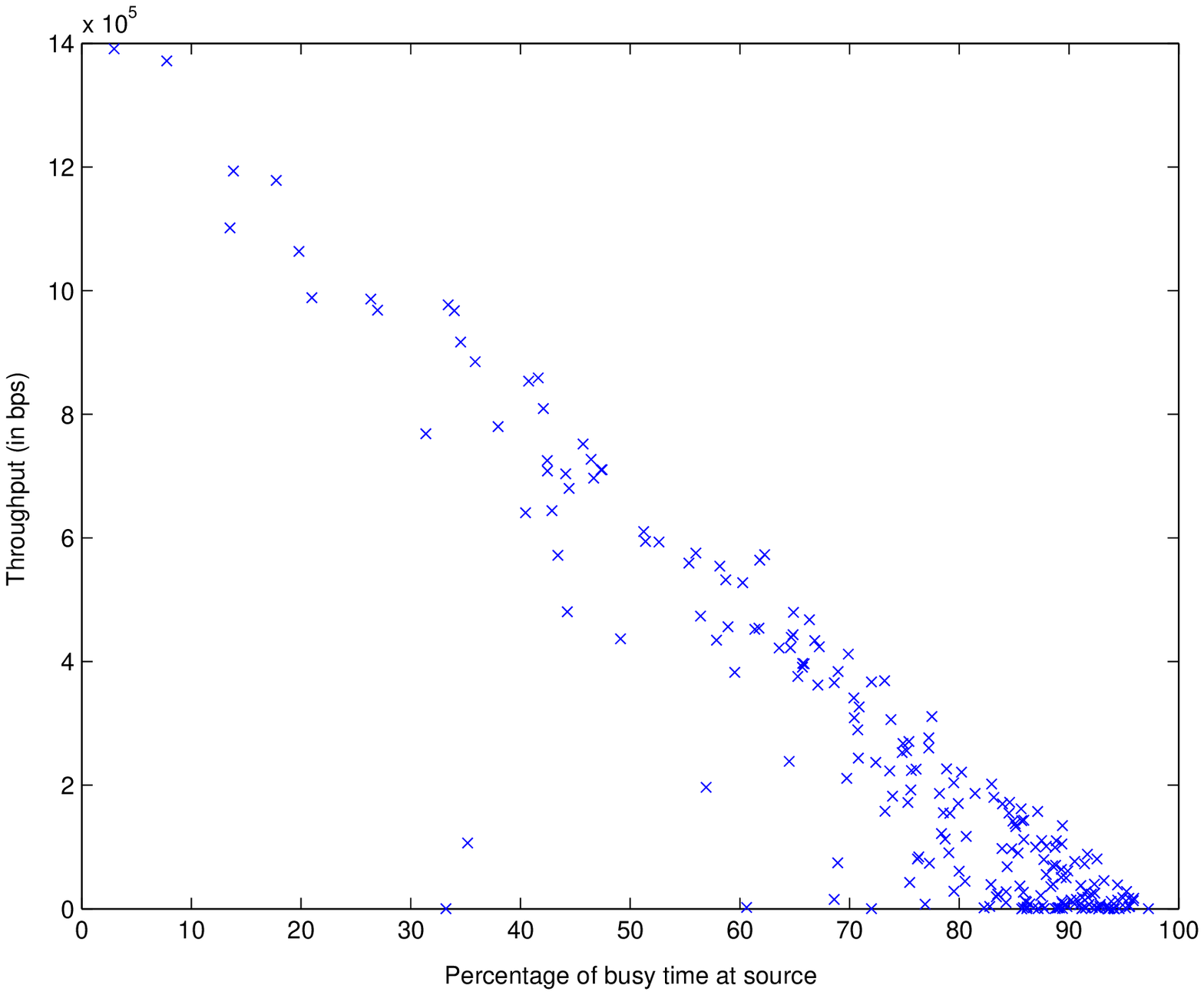}} 
         }
        \mbox{
        \subfigure[\label{fig:throughSimBusySrcHirateGt70}]{\includegraphics[scale=0.25]{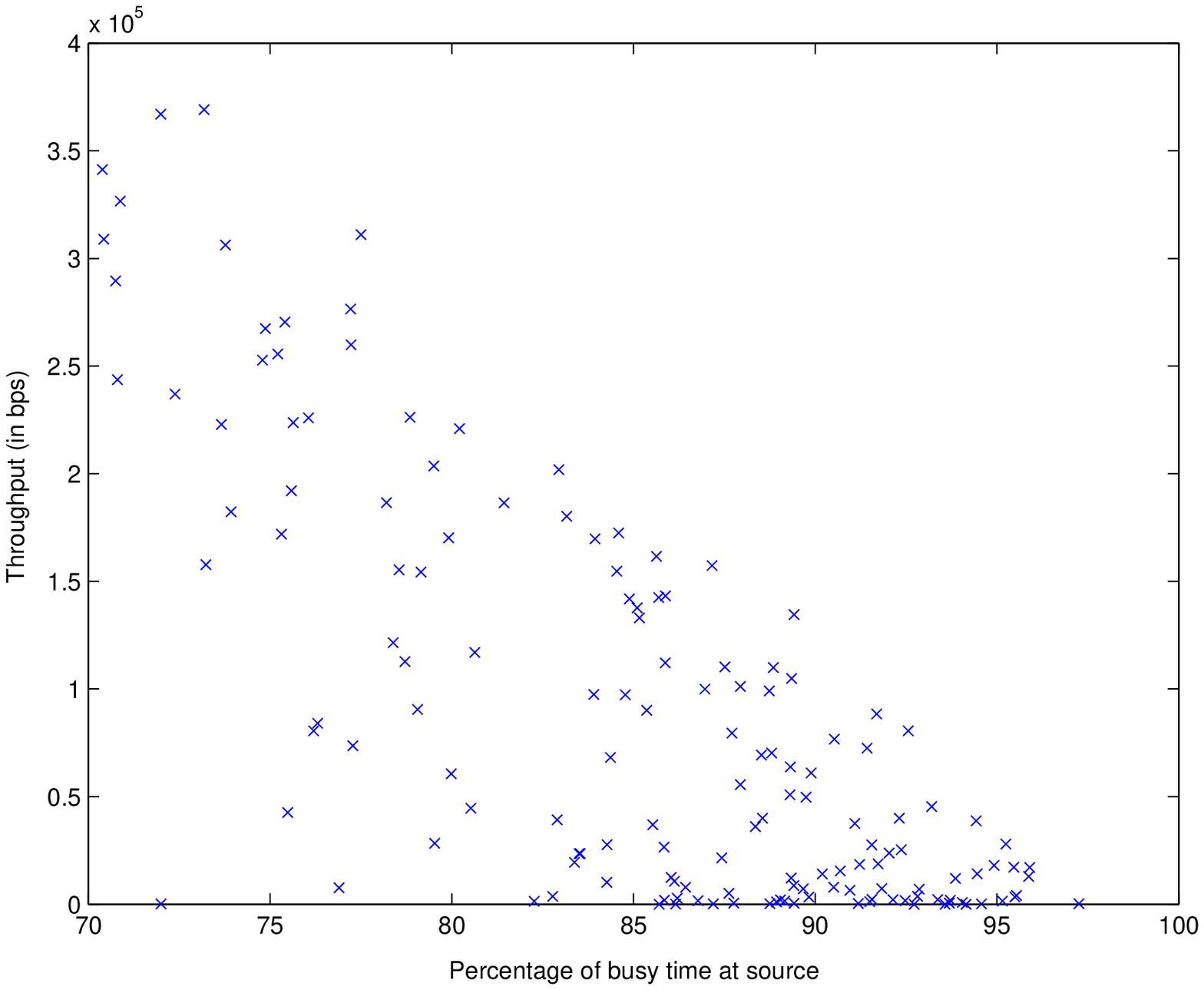}} 
        }
        \mbox{
        \subfigure[\label{fig:throughPerSimBusySrcCollsHirate}]{\includegraphics[scale=0.25]{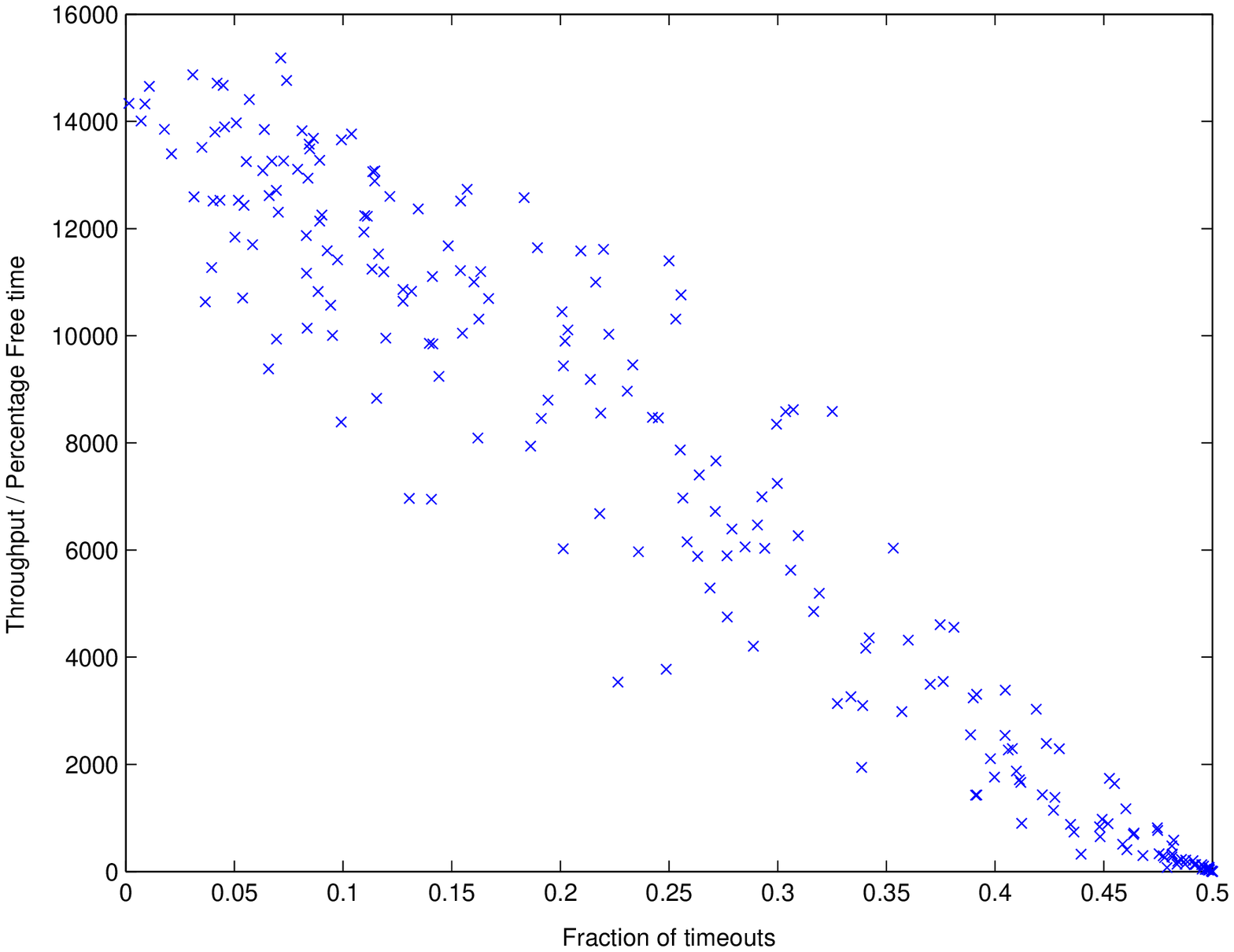}} 
        }
%       \mbox{
%         \subfigure[\label{fig:throughCollsHirateGt70}]{\includegraphics[scale=0.25]{figs/throughCollsHirateGt70.eps}} 
%         }
%       \mbox{
%         \subfigure[\label{fig:busySrcCollsHirateGt70}]{\includegraphics[scale=0.25]{figs/busySrcCollsHirateGt70.eps}} 
%         }
\caption{Effect of Interference and Scheduling on Capacity. (a) shows
         effect of source busy time on throughput; (b) enlarges the
         high interference region of (a); (c) shows that normalized
         throughput is a function of scheduling/MAC level timeouts.}
\end{center}
\label{fig:through}
\end{figure*}

The amount of time a source observes a busy channel is a commonly used
estimate of the capacity available to the link. For example, if the
MAC uses carrier sense, the sender will not send if the medium is
busy.  Intuitively, under an ideal scheduler, the throughput of the
link will only be a function of the \textit{available transmission
  time (ATT)}. However, in practical CSMA/CA based schedulers, packet
collisions and timeouts arise because perfect scheduling based on
local information only is not possible.  For example, the channel may
be sensed idle, at the source, but a collision occurs at the
destination.  Thus, the observed throughput will be a function of the
available capacity as well as the scheduling interactions.

% Existing MHWN modeling studies (e.g.,
% ~\cite{ref:jain_03_interference,ref:kodialam_03_routeSchedule,ref:kolar_06_mcf})
% and many protocol efforts (e.g., ~\cite{ref:ahn_02_swan}) use local
% estimates of interference to estimate capacity.  In this section, we
% motivate our work by demonstrating that using interference alone is
% insufficient for predicting the throughput of the connections. 
Figure \ref{fig:throughSimBusySrcHirate} plots the busy time\footnote{
  We experimented with other metrics such as busy time at destination
  and SINR at both the source and destination with very similar
  results.}  against the observed throughput of the link.  In general,
as interference increases, the achievable capacity decreases.
However, it can be seen that at higher busy times (enlarged for
clarity in Figure~\ref{fig:throughSimBusySrcHirateGt70}), large
variations in throughput arise for the same observed busy time.
% This
% effect can be seen more readily in Figure
% \ref{fig:throughSimBusySrcHirateGt70} which is an enlarged plot of the
% high busy time region of Figure \ref{fig:throughSimBusySrcHirate}.
% Clearly, the busy time poorly predicts the achievable capacity of
% links in this region.

With increasing interference, the scheduling effectiveness starts to
play a determining effect on the throughput of the link.  Let $t_i$ be
the throughput achieved by an ideal scheduler; intuitively $t_i$ is
proportional to the amount of {\em available transmission time (ATT)}.
Let $t_o$ be the observed throughput in the CSMA based scheduler, IEEE
802.11.  The ratio of $\frac{t_o}{t_i}$, called {\em normalized
  throughput} provides a measure of the scheduling efficiency relative
to an ideal scheduler independently of the available transmission
time.  Figure \ref{fig:throughPerSimBusySrcCollsHirate} plots the
normalized throughput as a function of observed percentage of MAC
level transmissions that experience RTS or ACK timeouts for
\textit{all} the links.  It can be seen that as the fraction of packet
timeouts increases, the normalized throughput decreases almost
linearly.  Thus {\em the reason for variations in observed capacity
  from the nominal capacity predicted by the interference metric is
  the scheduling as observed in MAC level timeouts.}
% We show later that the scheduling effects are a function of the
% relationships (or location under idealized propagation assumptions) of
% contending sources to each other.  More specifically, these
% relationships determine whether contending links are able to arbitrate
% the medium successfully using carrier sense and/or collision avoidance
% or whether uncontrolled collisions will occur resulting in high
% percentage of MAC level timeouts. 

Aggregate interference metrics such as busy time cannot predict
scheduling effects and correlate 
poorly with scheduling efficiency.  As a result, in the next section,
we analyze the problem of estimating the scheduling effects that a set
of active links present to each other to develop link quality metrics
capable of capturing both the interference and scheduling effects.

\section{Modeling Scheduling Interactions}
\label{sec:schedFormulation}

% Aggregate measures of ink quality were shown to be
% for the poor correlation is that those measures do not consider the
% subtle low-level interactions that occur during medium access.  In
% this section, we analyze these interactions and formulate link quality
% metrics that account for them.

In this section, we develop models for estimating the effect of
scheduling on the capacity of a link.  A key observation that provides the
starting point for this analysis is that the vast majority of
collisions occur due to interfering links whose sources are not
prevented from concurrent transmission by the MAC protocol; other
interfering links that can handshake effectively through the MAC
protocol are not impacted negatively by the scheduling effects.  
% The
% remaining collisions are due truly concurrent transmissions, which
% represent a race condition where the MAC protocol does not have a
% chance to mediate; as we show later in Section
% \ref{sec:micsValidation}, the effect of these collisions is small
% (less than 3\% of collisions in the scenarios we studied) and we do
% not consider them further in our analysis.

In contention based protocols, the channel state at the receiver is
not known at the sender which gives rise to the well-known hidden and
exposed terminal problems~\cite{ref:bhargavan_98_macaw}.  To counter
these effects, IEEE 802.11 uses an aggressive CSMA to attempt to
prevent far-away interfering sources from transmitting together (but
potentially preventing non-interfering sources from transmitting;
hidden terminal is reduced, but exposed terminal increased).  Despite
this aggressive carrier sense value, not all potential hidden
terminals are prevented.
Optionally, the standard allows the use Collision Avoidance (CA),
which consists of Request-to-Send (RTS)/Clear-to-Send (CTS) control
packets, to attempt to reserve the medium.  Thus, contention is
carried out with small packets.
% Specifically, the sender sends an RTS packet including the length of
% the transmission.  Neighbors that receive it defer from transmission
% for this duration; a process called Virtual Carrier Sense(VCS).  The
% receiver upon receiving the RTS packet responds with a CTS if the
% channel is clear at its end; the CTS informs its own neighbors of the
% duration of the transmission as well. 
However, CA only prevents interfering nodes in reception range of each
other (a small subset of possible interferers) and is often not
enabled in real deployments.  Thus, neither CSMA nor CA block prevent
all potential collisions, which gives rise to the destructive
scheduling interactions being studied in this paper.  We note that
many radios turn off the CA mechanism, increasing collision potential
and the effect of scheduling.  In this paper, we model the case with
CA enabled; modeling the case with CA disabled is simpler and similar
to the ACK timeout portion of the model.

\subsection{Preliminaries}

Let $G(V,E)$ be a graph representing the network where $V$ is the set
of all the nodes and $E$ is the set of active links.  Let
$\Theta_{ij}$ be a $n$x$n$ matrix, representing the signal
strength observed at node $j$ for node $i$'s transmission.  Under
ideal power-law signal propagation assumptions, $\Theta_{ij}$ can be
derived based on node location.  However, it may be defined according
to other propagation models or experimentally derived based on
observed connectivity and interference.

We show the derivation assuming that the RTS/CTS handshake
% \footnote{An
%   RTS transmission informs neighbors of an upcoming data transmission
%   and its duration.  If the medium is available at the receiver, it
%   responds with a CTS that tells its own neighbors the same.  Finally,
%   data is sent and an acknowledgement sent is in response if the data
%   is received correctly by the receiver.} 
is enabled (which is the more difficult case).  However, it is
possible to investigate performance without this handshake since
disabling the handshake is becoming more common in
practice~\cite{ref:xu_02_interference}.  We showed in
Section~\ref{sec:correlation} that collisions accurately predict the
effect of scheduling on capacity; they are the primary indicators of
inefficient scheduling. In addition to wasted channel time, they
result in back-offs, which may cause the channel to become idle.  

At the source, collisions are observed in one of two ways: an RTS
Timeout (indicating a loss of RTS or absence/loss of CTS) or a DATA
timeout (indicating a loss of DATA or ACK).  The root cause of
collisions is the presence of sources that can transmit concurrently
but that interfere with each other's destinations; ideally the MAC
protocol would have prevented all but the non-interfering sources from
transmission.  We break the problem into two parts: (1) identifying
the links whose senders can transmit concurrently despite the
protocol's handshake; and (2) Quantifying how these
links interact.

\subsection{Identifying Concurrently Transmitting Senders}
\label{sec:mics}
We define the notion of a \emph{Maximal
  Independent Contention Set (MICS)} as a set of links whose sources
are not prevented from initiating transmissions while there are active
transmissions on the other links of the same MICS\footnote{This definition bears similarities to maximum cliques (e.g.,~\cite{ref:garetto_06_starve}) with the exception that it is restricted to active links.}.  
Due to the imperfect nature of MAC handshake, links that
interfere may belong in the same MICS.  Thus, {\em collisions occur
  due to interactions within an MICS (except for rare race conditions),
  transmissions across different MICS' are prevented by the MAC
  protocol}.
%%  (with the exception of instantaneous
%%   transmissions that arise due to two sources sensing the medium to be
%%   idle and transmitting; this effect is minor in a
%%   randomized protocol).  \rednote{Vinay, can we make this argument
%%   quantitatively (how many of the collisions are due to this effect
%%   vs. the other effects)}\bluenote{Vinay: Nael, we can do it. Will try
%%   to get it done for Mobicom. Transmission time of RTS is of the order
%%   of nano seconds and there should not be many such drops (compared to
%%   the ones that are dropped often because of these effects)}
The definition of an MICS follows.
\begin{definition}
\label{def:mics}
A set $C \subseteq E$ forms a \emph{Maximal Independent Contention
  Set (MICS)} if 
$\forall (s,d) \in C, \nexists (s_2,d_2) \in E-C \text{ such that }\\
  \Theta_{s_2d} + \displaystyle \sum_{(s_1,d_1) \in C, (s_1,d_1) \neq (s,d)}
  \Theta_{s_1d} < \rxSen - W$. \\A set of all MICS is denoted by
  $\mis$. 
\end{definition}
\noindent
A set of all the MICS in which a given link $(s,d)$ is present is
denoted by $\edgePresence_{(s,d)}$ (Equation \ref{eqn:edgePresence}).
%as given in Equation \ref{eqn:edgePresence}.
%\begin{eqnarray}
%\label{eqn:edgePresence}{\edgePresence}_{(s,d)} = \{C | C \in \mis, (s,d) \in C\}\\
%\end{eqnarray}

\begin{eqnarray}
  \label{eqn:edgePresence}{\edgePresence}_{(s,d)} = \{C | C \in \mis, (s,d) \in C\}
\end{eqnarray}
\noindent The probability that MICS $C$ is currently active is
represented as ${\probMis}_C$. Precise formulation of ${\probMis}_C$
requires estimation of different factors like: (1) the dependence
of MICS on the initiation order of sources; and (2) the sequence of
interactions between the conflicting sources (probably cyclic).
% (1) the number of MICS in which 
% each source belongs to; (2) the dependence of the MICS on the order in 
% which the sources of all the links initiate; 
% (3) information about chains of such interactions among conflicting sources
% (probably cyclic) (4) the set of MICS that can be activated given that a 
% subset of the links are already active. 
An approximation of ${\probMis}_C$ is desirable since an accurate
estimation of all the factors makes the model computationally
infeasible.  We use the approximation in Equation \ref{eqn:probMis}
which assumes that the set of sources capable of initiating concurrent transmissions
compete independently from each other. The probability of a MICS occurring is
dependent upon the number of sources present in the MICS. Thus,
the approximation makes all MICS' of the same size equi-probable, with
the probability of a MICS increasing with the number of edges in it.
\begin{eqnarray}
  \label{eqn:probMis} \probMis_{C} = \frac{|C|}{\sum _{C' \in \mis}|C'|}
\end{eqnarray}
\noindent
Improving the characterization of MICS activation probabilities is an
area of further refinement of the model.

% Computing this probability precisely is a hard problem and
% requires detailed analysis of the order sources who compete, 
% their conflicting links, the number of MICS each source belongs to and 
% information about chains of such interactions among conflicting sources
% (probably cyclic).  Instead, we
% approximate it by assuming that the set of sources compete
% independently from each other (Equation~\ref{eqn:probMis}).  
% Thus, the amount of time a set of concurrent links is scheduled is proportional
% to the number of links present in such a set.
% Thus,
% the approximation makes all MICS' of the same size equi-probable, with
% the probability of a MICS increasing with the number of edges in it.
% \bluenote{TODO: Give example here}
% Thus, a higher number of edges indicates a higher probability of the MICS 
% winning the contention.
%The probability of the occurrence of MICS $C$ is represented as
%${\probMis}_C$. Assuming the set of sources compete independently from each other, the probability of an occurance of each MICS is dependent upon the number of sources in the MICS. The overall probability of occurance of MICS $C$ is represented by ${\probMis}_C$ and can be approximated by Equation \ref{eqn:probMis}
% as given in Equation \ref{eqn:probMis}.
% Even though MICS represent the source view of scheduling initiation
% and the contention among multiple sources are not independent, it
% provides a light-weight first order approximation for arriving at a
% single rating for the link. 

\subsection{Quantifying Link Interactions}

The majority of the timeouts occur due the interaction among the
sources which do not prevent each other from transmission via CSMA or
CA.  MICS characterize such groups of sources.  Evaluating link
interactions is therefore simplified into evaluating interactions on a
per-MICS basis and then weighting these by the probability of MICS
activation in Equation~\ref{eqn:probMis}. The overall timeouts
experienced by a link is a function of the the packet timeouts due to
interaction in all the MICS in which the link is present. In this
section, we model the problem at two levels: (1) Interactions between
the links of the MICS; and (2) A weighting function across MICS to
measure the overall timeouts. %  In this study, we estimate the RTS
% timeouts and the ACK timeouts for the IEEE 802.11. 
While we use the IEEE 802.11, the analysis can be adapted to other
contention based protocols.
% The generic
% framework of the model can be extended for other protocols and
% variations of IEEE 802.11.

\paragraph{Estimating RTS Timeouts}
An RTS timeout occurs for one of three reasons: (1) an RTS collision;
(2) a busy receiver not responding to the RTS due to physical or
virtual carrier sensing being active at the destination; or (3) a CTS
collision.

Consider an RTS timeout at a link $(s,d)$. Since the noise at $s$ is
below $\rxSen$ before the RTS initiation (otherwise, the link is not
part of the currently active MICS) and $d$ should respond with CTS
immediately upon reception of RTS, the \textit{chances of CTS
  collision are low}.  This intuition was validated by extended
observation of MAC traces in simulation.  Hence, an RTS collision or a
busy channel at the receiver are the candidates for causing an RTS
timeout.

% 
% Consider a more complicated scenario (4). If the channel is busy at 
% receiver $d$, then it will be same as scenario (2). But, if the 
% channel at $d$ is idle and VCS is set, then the VCS should have been 
% set by overhearing a (a)foreign-CTS; or (b) a foreign-RTS which has 
% timed out (and the VCS still assumes ongoing transmission). 
% Under case (b), the timeout of the current RTS to $d$ by $s$ would 
% trigger a sequence of such `False VCS' to neighbors of the source $s$. 
% Such `False VCS' cause chains of interaction is currently not 
% addressed in our study. We would like to address them in future 
% extentions using a probabilistic approach. 
% 
% Hence, an RTS collision or a busy channel at the receiver are the 
% candidates for causing an RTS timeout when VCS is not considered.

In a given MICS $C$, an \textit{Unsafe Link} $(s_1,d_1)$ for a link
$(s,d)$ is given by Equation \ref{eqn:edgeUnsafe} and is defined as a
link whose transmission at source ($s_1$) may create a busy channel at
the receiver $d$ (condition 1 of Equation \ref{eqn:edgeUnsafe}) or
cause an RTS collision (condition 2). Both the cases results in an RTS
Timeout.
\begin{eqnarray}
\nonumber \label{eqn:edgeUnsafe}{\edgeUnsafe}^{C}_{(s,d)}=\Bigl \{(s_1,d_1) \mid
(s_1,d_1) \in C, \\
\Theta_{s_1d} \ge \rxSen - W \text{ or } \frac{\Theta_{sd}}{\Theta_{s_1d} + W} < \sinrThresh \Bigr \}
\end{eqnarray}

\bluenote{Vinay: Nael, the below paragraph gives one of our
  limitations. Does the it sound right?}  A limitation of the current
formulation is that we do not consider the effect of timeouts due to
the virtual carrier sense being set at the receiver.  While a
substantial number of RTS timeouts were due to the VCS (around 25\% as
explained later in \ref{sec:micsValidation}), modeling this effect
requires estimating: (1) Packet capture ability of the node under a
given MICS (which cannot be assumed to be a constant ``reception
threshold'' as done in Protocol Model of interference); (2)
Distinguishing the timeouts due to ``False VCS'' which may turn on the
VCS even if the competing packet transfer is unsuccessful (that is, an
RTS may cause a sender to defer, even if its CTS was never received --
we call this ``false VCS'' effect).  A deeper study of VCS related
effects is an area of extension of the model.

The number of RTS Timeouts for a link $(s,d)$ depends upon the number
of unsafe links that can be initiated before the link $(s,d)$. Deriving 
the probabilities of such initiations is done as follows:
Let $p_e(k)$ be the probability that \textit{exactly} $k$ links will initiate
transmission before a given link. It can be shown that this
probability is given by $p_e(k)=\frac{1}{k+1}$. Let $p(k)$ be the probability 
that at least one of the $k$ links will initiate transmission before any 
given link. It can be shown that 
$p(k) = \sum_{i=1}^{k} {(-1)}^{(i-1)} {k \choose i} p_e(k)$. 
Given $k$ unsafe links, $p(k)$ indicates the probability that at least
one of the unsafe link has initiated before the link $(s,d)$.
A cumulative metric to measure the percentage of the RTS Timeouts should
be unbiased to the number of MICS to which the link belongs. We
approximate the such a metric by ${\rtsRating}_{s,d}$.
\begin{definition}
  We define the following \textit{Interaction Based RTS Link Rating}
  ($\rtsRating_{(s,d)}$) for a link
  $(s,d)$:\\
  ${\rtsRating}_{(s,d)} = \dfrac{\displaystyle \sum_{C \in \mis}
    p(|{\edgeUnsafe}^{C}_{(s,d)}|)}{|{\edgePresence}_{(s,d)}|}$
\end{definition}
% 
% The number of RTS timeouts depends upon the number of unsafe links
% ($\edgeUnsafe^C_{(s,d)}$).  Since the `percentage' of timeouts is
% desired, the rating should be unbiased towards the number of MICS the
% given link belongs. The above definition would approximate the
% percentage of the RTS Timeouts. 
% The number of RTS timeouts depends upon the number of unsafe links($\edgeUnsafe^C_{(s,d)}$). 
% Since the `percentage' of timeouts is desired, the rating should be unbiased towards the 
% number of MICS the given link belongs. The above definition would approximate the percentage
% of the RTS Timeouts.
%
% The number of RTS timeouts for a given link $(s,d)$ under a given MICS $C$ is dependent upon the count of 
% the edges in the MICS that cause the RTS timeouts($\edgeUnsafe^C_{(s,d)}$) and the probability of their 
% activation before the RTS of $(s,d)$ is transmitted. The cumulative effect for all MICS is captured by the 
% numerator. Since we are interested in the percentage of the drops, the bias towards the nodes that belongs 
% to a larger number of MICS is eliminated by the denominator.
%We show in Section \ref{sec:schedRes} that the above RTS Timeout rating strongly matches with the observed percentage of RTS Timeouts.

\paragraph{Estimating ACK Timeouts}

More important than RTS timeouts are collisions affecting DATA or ACK
packets which result in ACK timeouts. ACK timeouts are costly
since they denote unsuccessful communication after a prolonged period
of channel usage.
% The
% assessment of ACK timeout is challenging since approximation is harder
% because there are more possibilities of DATA packet getting corrupted
% due to the extended duration of transmission.
%It should even account for the fact that the start of DATA
%transmission happens only after an initial handshake is
%successful. Since the CCA is not done before ACK transmission, the
%ACK will be sent irrespective of the channel state at receiver.  
Considering that the ACK will be sent immediately after DATA
transmission, the probability of the ACK collision at the source is
small. Hence we formulate the ACK timeouts due to DATA packet
corruption.
% We also verify the reasoning by ACK timeouts by from simulator
% results.
The basis for modeling ACK timeouts is to determine the links that can
corrupt an ongoing DATA packet by RTS, CTS or ACK packet.
We define a set of links that can corrupt the DATA transmission of the
link $(s,d)$ by initiating an RTS by $\RTSCorrupt_{(s,d)}$ as given
in 
Equation \ref{eqn:RTSCorrupt}. These are the set of links $(s_1,d_1)$
which co-exist in any of the MICS of the link and the RTS from $s_1$
could corrupt the data packet at $d$.
\begin{eqnarray}
\nonumber \label{eqn:RTSCorrupt} \RTSCorrupt_{(s,d)} = \Bigl \{(s_1,d_1) \mid \exists C \in \mis \text{ such that } \\
(s,d),(s_1,d_1) \in C, \frac{\Theta_{sd}}{\Theta_{s_1,d} + W} < \sinrThresh \Bigl \}
\end{eqnarray}

Similar to $\RTSCorrupt$, the
sets of links that can corrupt the DATA packet by transmitting CTS or
ACK are represented as $\CTSCorrupt_{(s,d)}$ and $\ACKCorrupt_{(s,d)}$
respectively. To determine the set $\CTSCorrupt_{(s,d)}$, we place the
restriction that RTS does not corrupt the data packet to ensure that
the same DATA packet drop is not accounted twice, in
$\RTSCorrupt_{(s,d)}$ and $\CTSCorrupt_{(s,d)}$. Equation \ref{eqn:CTSCorrupt}
shows the set of links that can corrupt the packet of the link $(s,d)$
by CTS transmission.
\begin{eqnarray}
\nonumber \CTSCorrupt_{(s,d)} = \{(s_1,d_1) \mid \exists C \in \mis \text{ such that } \\
\nonumber (s,d),(s_1,d_1) \in C, (s_1,d_1) \not \in \RTSCorrupt_{(s,d)}, \\
\label{eqn:CTSCorrupt} \Theta_{sd_1} < \rxSen - N, \frac{\Theta_{sd}}{\Theta_{d_1d} + W} < \sinrThresh\}
\end{eqnarray}

Similarly, the set of links that can corrupt the DATA packet by ACK
transmission are defined in equation \ref{eqn:ACKCorrupt}. There may
be links whose CTS does not corrupt the DATA packet of but the ACK
packet could corrupt since ACK transmission $\cal{T}$ 
A $\ACKCorrupt_{(s,d)}$ for a given link is a set of links
$(s_1,d_1)$ such that the links co-exist in the same MICS and
$(s_1,d_1)$ does not belong to either $\RTSCorrupt_{(s,d)}$ or
$\CTSCorrupt_{(s,d)}$ and an RTS from either $(s,d)$ or $(s_1,d_1)$
does not corrupt the DATA packet of the other link, while the ACK
packet from $d_1$ corrupts the DATA of $(s,d)$. There may be links
$(s_2,d_2)$ whose CTS does not corrupt the DATA packet of $(s,d)$
because CTS initiation is canceled because of the channel being busy
at $d_2$. However, if the scheduling of two links is such that
$(s_2,d_2)$ starts before $(s,d)$ and $(s_2,d_2)$ ends before $(s,d)$
then the ACK packet of $d_2$ may corrupt the DATA packet at $d$ since
ACK transmission happens without CCA.  
\begin{eqnarray}
\nonumber \ACKCorrupt_{(s,d)} = \{(s_1,d_1) \mid \exists C \in \mis \text{ such that } (s,d),(s_1,d_1) \in C, \\
\nonumber (s_1,d_1) \not \in \RTSCorrupt_{(s,d)}, (s_1,d_1) \not \in \CTSCorrupt_{(s,d)}, \\
\label{eqn:ACKCorrupt} \frac{\Theta_{s_1d_1}}{\Theta_{sd_1} + W} >= \sinrThresh, \frac{\Theta_{sd}}{\Theta_{d_1d} + W} < \sinrThresh\}
\end{eqnarray}

While calculating the DATA drops, it is important to account for the
cumulative interference produced by multiple concurrent links. This is
because of the higher possibility of such combination of interference
during the longer period of data transmission. 
Let $\cumNoiseFactor_C^{(s,d)}$
be the probability that the interference from the links of the MICS $C$ corrupts 
the DATA packet. Calculation of $\cumNoiseFactor_C^{(s,d)}$ requires the estimation
of the interference caused to a node by multiple active links. In the next paragraph,
we explain the estimation of the $\cumNoiseFactor_C^{(s,d)}$ by capturing the 
cumulative interference.

Indirect Interference Edges $\indirectIferEdges_C^{(s,d)}$ for a given edge $(s,d)$ in a
given MICS $C$ is a set of links in $C$ whose transmission does \textit{not}
make the channel busy at $d$ (Equation \ref{eqn:indirectIferEdges}).
The maximum cumulative noise from indirect edges is the sum of maximum
noises observed at $d$ by either $s_1$ or $d_1$(Equation
\ref{eqn:maxCumNoiseBelowSen}). Simultaneous DATA transmissions by
Indirect Interference Edges may create a busy channel.
\begin{eqnarray}
\nonumber \label{eqn:indirectIferEdges}
\indirectIferEdges_C^{(s,d)} = \Bigl \{(s_1,d_1) \mid (s_1,d_1) \in C, (s_1,d_1) \ne (s,d), \\
\Theta_{s_1d} + W < \rxSen , \Theta_{d_1d} + W < \rxSen \Bigl \}
\end{eqnarray}
\begin{eqnarray}
\label{eqn:maxCumNoiseBelowSen} \maxCumNoiseBelowSen_{C}^{(s,d)} = \sum _{\forall (s_1,d_1) \in \indirectIferEdges_C^{(s,d)}} \mymax(\Theta_{s_1d}, \Theta_{d_1d})
\end{eqnarray}
The cumulative interference factor (Equation \ref{eqn:cumNoiseFactor})
defines the effect on data transmission because of such combination
of interference. If the interference $\maxCumNoiseBelowSen^C_{(s,d)}$ does not
corrupt the data packet, then the value $\cumNoiseFactor^C_{(s,d)}$ is
set to $0$. If it affects it then it is set to a value based on
probability of MICS $C$ and the number of indirect interfering
links (Equation \ref{eqn:cumNoiseFactor}). 
\begin{equation}
\label{eqn:cumNoiseFactor}
\cumNoiseFactor_C^{(s,d)} = \begin{cases}
        \probMis_C p_e(|\indirectIferEdges_C^{(s,d)}|) &  \text{if } \frac{\Theta_{sd}}{(\maxCumNoiseBelowSen_C^{(s,d)} + W)} < \sinrThresh\\
        0,      & \textit{otherwise.}
      \end{cases}
\end{equation}

Finally, an estimate for the amount of ACK Timeouts is obtained by
accounting for the links that corrupt DATA packet by RTS, CTS, ACK in
addition to indirectly interfering links as given in Equation
\ref{eqn:ackTimeoutRating}. Since, the percentage of timeouts is of
interest, the probability that the interfering links will corrupt the
data packet also depends upon ratio of the probability of the source
winning in a given MICS. The probability of the source of the link $(s,d)$ 
winning the channel is denoted by $\probSrcWin_{(s,d)}$ and is given by the
sum of all the probability of the MICS winning the channel in which the link
is present.
\begin{eqnarray}
\label{eqn:ackTimeoutRating} \dataRating_{(s,d)} = \dfrac{\displaystyle \sum_{\forall C \in \mis} \probMis_{C}.p(|\RTSCorrupt_{C}| + |\CTSCorrupt_{C}| + |\ACKCorrupt_{C}|) + \cumNoiseFactor_C^{(s,d)} }{\probSrcWin_{(s,d)}}
\end{eqnarray}

\paragraph{Impact of Virtual Carrier Sense on ACK Timeouts}
\label{sec:VCSImpact}
The formulation presented thusfar does not account for the effect of
Virtual Carrier Sensing (VCS).  More specifically, when an RTS or CTS
is received by other nodes, these nodes defer from
transmission even if the RTS or CTS ends up failing, meaning that the
data transmission will not occur. Estimating VCS effect requires
determination if a node can capture the RTS/CTS packets based on the
signal to interference at the node. Under the SINR propagation model, the 
packet can be captured if the cumulative interference due to all the 
concurrent active links yields a SINR ratio greater than the 
$\sinrThresh$.

In modeling this effect, and as a first order approximation, we
ignore the possibility of simultaneous RTS or CTS packets 
(as observed in Section \ref{sec:micsValidation}) and consider
only the DATA packets whose transmission causes extended interference
time. The cumulative interference in a MICS cannot be calculated
by summing up the interference experienced from all the sources of the MICS,
since MICS includes the links which may suffer RTS Timeouts, thus not 
initiating the DATA transfer. Measuring cumulative interference due
to concurrent DATA packet transmission can be done by computing the 
maximum independent set of links that can initiate the DATA transfer 
concurrently. We call such a set of links as \textit{Existent MICS} (EMICS).
\begin{definition}
\label{def:existentMics}
A set $C' \subseteq E$ is an \emph{Existent Maximal Independent Contention Set(EMICS)} if:
\begin{enumerate}
\item $\exists C \in \mis \text{ such that } C' \subseteq C\\$ 
\item $\forall (s,d) \in C' \text{ and } \forall (s_1,d_1) \in C'\\
\displaystyle \sum_{(s_1,d_1) \ne (s,d)} \mymax(\Theta_{s_1d}, \Theta_{d_1d}) + W < \rxSen\\$
\item $\forall (s,d) \in C' \text{ and } \forall (s_1,d_1) \in C'\\
\frac{\Theta_{sd}}{\displaystyle \sum_{(s_1,d_1) \ne (s,d)} \mymax(\Theta_{s_1d}, \Theta_{d_1d}) + W} \ge \sinrThresh\\$
\item $\forall (s,d) \in C' \text{ and } \forall (s_1,d_1) \in C'\\
\frac{\Theta_{ds}}{\displaystyle \sum_{(s_1,d_1) \ne (s,d)} \mymax(\Theta_{s_1s}, \Theta_{d_1s}) + W} \ge \sinrThresh\\$
\end{enumerate}
A set of all EMICS is denoted by $\optiMis$. 
\end{definition}

A link $(s,d)$ can be present in more than one EMICS, thus, the interference 
experienced at the node $a$ when the link $(s,d)$ is active will vary depending 
upon which EMICS is active. The maximum interference at the node $a$ when $(s,d)$ 
is active is the maximum of the sum of interference caused by links in EMICS of 
$(s_1,d_1)$. (Equation \ref{eqn:maxParallel}).
\begin{eqnarray}
% Maximum Noise at 'a' given that (s,d) is active
\label{eqn:maxParallel} \maxParallel_{a,(s,d)} = \displaystyle \max_{\forall C' \in \optiMis, (s,d) \in C'} \Biggl(\displaystyle \sum _{(s_1,d_1) \in C'}\Theta_{as_1} \Biggr)
\end{eqnarray}

We approximate that the node $a$ can set its VCS to the RTS or CTS of
link $(s,d)$ if the noise at $a$ is lesser than
$\maxParallel_{a,(s,d)}$. This is given by $\navSensing_{a,(s,d)}$ in
Equation \ref{eqn:navSensing}.
\begin{equation}
\label{eqn:navSensing}
% Nav sensing at s from either s_1 or d_1. Would s hear to RTS or CTS of (s_1,d_1)
\navSensing_{a,(s,d)} = \begin{cases}
      1 &  \text{if } \mymax(\Theta_{as}, \Theta_{ad}) > \rxSen \text{ and } \\~&~~~\frac{\mymax(\Theta_{as}, \Theta_{ad})}{\maxParallel_{a,(s,d)}} \ge \sinrThresh\\
      0,      & \textit{otherwise.}
      \end{cases}
\end{equation}

% The VCS at node $a$ for the ongoing
% transmission $(s,d)$ will be set if the interference experienced does not corrupt
% the RTS or CTS packet from the link $(s,d)$. 
% Since the interference varies based on 
% different active EMICS, a single value estimate for the packet capture for VCS is not
% possible. As a first order approximation, the maximum cumulative interference 
% experienced from the EMICS in which the link $(s,d)$ is present is considered to 
% decide if VCS can be set at node $a$.

% \bluenote{Vinay:Nael, I dint understand this. Commented and rewrote}The 
% set of all EMICS is denoted as $\optiMis$. By measuring the maximum 
% interference due to EMICS at a node while a given link is active, we can 
% account for the effect of VCS on $\RTSCorrupt$ and $\CTSCorrupt$. 
% Given that a link $(s,d)$ is active in an EMICS $C'$, the interference 
% caused at a node $a$ will be created by other sources of $C'$. The maximum
% interference(denoted by $\maxParallel_{a,(s,d)}$) created at $a$ when 
% $(s,d)$ is active can be captured by considering all the EMICS to which 
% $(s,d)$ belongs. As a first order approximation, $a$ can set its VCS to 
% packet of edge $(s,d)$ if the $\maxParallel_{a,(s,d)}$ observed cannot 
% corrupt the packet from $(s,d)$. 
% \rednote{Wouldn't this result in a pessimistic estimate of VCS as in
%   many cases not the whole EMICS will be active?} 
% \bluenote{Vinay: Same answer as for the above question. Most of the EMICS
% should go on together most of the time} 
% This
% formulation increases the runtime of the algorithm to $O(l_{\optiMis}
% m^3)$.

We redefine the set of links that can corrupt the data packet by
transmitting RTS or CTS based on the VCS and denote the sets by
$\RTSCorruptImp_C$ and $\CTSCorruptImp_C$ as given in Equations
\ref{eqn:RTSCorruptImp} and \ref{eqn:CTSCorruptImp}.
\begin{eqnarray}
\label{eqn:RTSCorruptImp} \RTSCorruptImp_{(s,d)} = \{(s_1,d_1) \mid (s_1,d_1) \in \RTSCorrupt_{(s,d)}, \navSensing_{s_1,(s,d)} = 0\} \\
\label{eqn:CTSCorruptImp} \CTSCorruptImp_{(s,d)} = \{(s_1,d_1) \mid (s_1,d_1) \in \CTSCorrupt_{(s,d)}, \navSensing_{d_1,(s,d)} = 0\}
\end{eqnarray}

The ACK Timeout Rating accounting for the VCS is given in Equation \ref{eqn:ackTimeoutRatingImp}. 
\begin{eqnarray}
\label{eqn:ackTimeoutRatingImp}\dataRatingImp_{(s,d)} = \dfrac{\displaystyle \sum_{\forall C \in \mis} \probMis_{C}.p(|\RTSCorruptImp_{C}| + |\CTSCorruptImp_{C}| + |\ACKCorrupt_{C}|)+ \cumNoiseFactor_C^{(s,d)}}{\probSrcWin_{(s,d)}} 
\end{eqnarray}
All the demonstrated results use VCS enabled ACK Timeout rating.

\paragraph{Complexity Analysis} 
Contention
among the links can be mapped into a \textit{Contention Graph}
$G(V',E')$ where the vertices represent the active links and the edges
denote the contention. An edge is present between two vertices $a,b
\in V'$ if the source's of $a$ and $b$ can initiate concurrent
transmission~\cite{ref:jain_03_interference} under contention based
scheduling. The computation of the set of MICS($\mis$) is the problem
of finding all the \textit{Maximal Independent Sets} (MIS') in $G'$
(all the Maximal Cliques in the complementary graph $\bar{G'}$). The
problem of finding \textit{MIS} is NP-complete. The number of MIS in
a graph is denoted by $l_{\mis}$, which is bounded by
$O(3^{\frac{n}{3}})$\cite{ref:moon_65_cliques}. We used an 
iterative heuristic \cite{ref:jain_03_interference} for finding
independent sets. The algorithm implemented has a complexity of
$O(l_{\mis}n^3)$ where $n=|V'|$, the number of active links. The
approximate algorithm without considering the effect of VCS (Section
\ref{sec:VCSImpact}) has $O(l_{\mis}n^2\log(n))$ complexity.
\subsection{Experimental Validation}
\label{sec:schedRes}
This section validates MICS interactions, which form the 
basis of the formulation, with the interactions observed in 
the simulation. We evaluate the ability of the IBLR metrics to
predict MAC collisions in Section \ref{sec:iblrMetrics}.
% Section \ref{sec:routingRes} compares the effectiveness of the 
% SAR scheme with the routes obtained via a conventional routing 
% algorithm (DSR) and those obtained from a coordinated routing 
% formulation that uses interference metric. 
\rednote{Should consider moving the IBLR analysis right after the IBLR derivation}
\subsubsection{Validation of MICS Interaction}
\label{sec:micsValidation}
% \begin{figure}
% \centering
% \includegraphics[scale=0.35]{figs/interMis.eps}
% \caption{Breakdown of Collision Causes}
% \label{fig:res.collMics}
% \end{figure} 
The link quality formulation results were analyzed and were compared
with the simulation results. 
We altered the QualNet
simulator~\cite{ref:qualnet} to account for the \textit{SINR
  Threshold} model. Ten different scenarios, each with a random 
placement of 144 nodes in a 1600m$\times$1600m area was chosen 
with 25 one-hop connections (250 active links).
%  This
% experiment was repeated for 10 different configurations (250 active
% links) and each configuration was run for 20 seeds.

Figure \ref{fig:res.collMics}
compares the simulation results of the causes for packet timeouts 
(X-axis) to the average number of instances on Y-axis. 
%The number of instances is the average of the
%above mentioned 10 random scenarios that were the observed in the
%simulation. 
\textit{RTS-A} and \textit{RTS-I} stands for the RTS packet collision 
during the \textit{arrival} and \textit{intermediate stage} of the RTS packet. 
% \textit{RTS-I} denotes the RTS
% packet collision during the \texttt{intermediate} stage of RTS packet
% reception. 
Similarly, \textit{DATA-A} and \textit{DATA-I} stands for
the DATA packet collision on arrival and at intermediate 
stage. The \textit{CTS-PCS} and \textit{CTS-VCS} denotes the CTS packet 
not being sent by the destination as a result of physical and virtual carrier
sensing , respectively.
% due to channel busy at the destination and CTS-VCS denotes that CTS not being sent due
% to VCS indication of ongoing transmissions at the receiver even though
% the channel is idle.
Intra-MICS interactions are those due to two transmissions of the same
MICS (which are captured by the formulation).  Inter-MICS denotes
transmissions that are not captured by the formulation and can be due
to random effects of channel access like instantaneous transmissions.
As we indicated earlier, Inter-MICS collisions are quite rare 
(2.67\% in our results)
with most occurring in the \textit{RTS-A} and \textit{RTS-I} stage. 
This validates the MICS framework for capturing the scheduling 
interactions. 
% It can be also observed that DATA packets generally
% experience collisions at intermediate reception stage and foreign
% transmissions conflicting with DATA packet are not initiated in between
% the CTS transmission and the DATA reception (since DATA-A is zero).
The intermediate DATA packet collisions indicate the ineffectiveness 
of the RTS-CTS handshake (including the VCS).

\subsubsection{Effectiveness of the IBLR metrics}
\label{sec:iblrMetrics}
\begin{figure*}
\begin{center}
%       \mbox{
%       \subfigure[Breakdown of Collision Causes\label{fig:res.collMics}]{\includegraphics[scale=0.27]{figs/interMis.eps}} \quad
%       }
        \mbox{
        \subfigure[Collision Causes\label{fig:res.collMics}]{\includegraphics[scale=0.25]{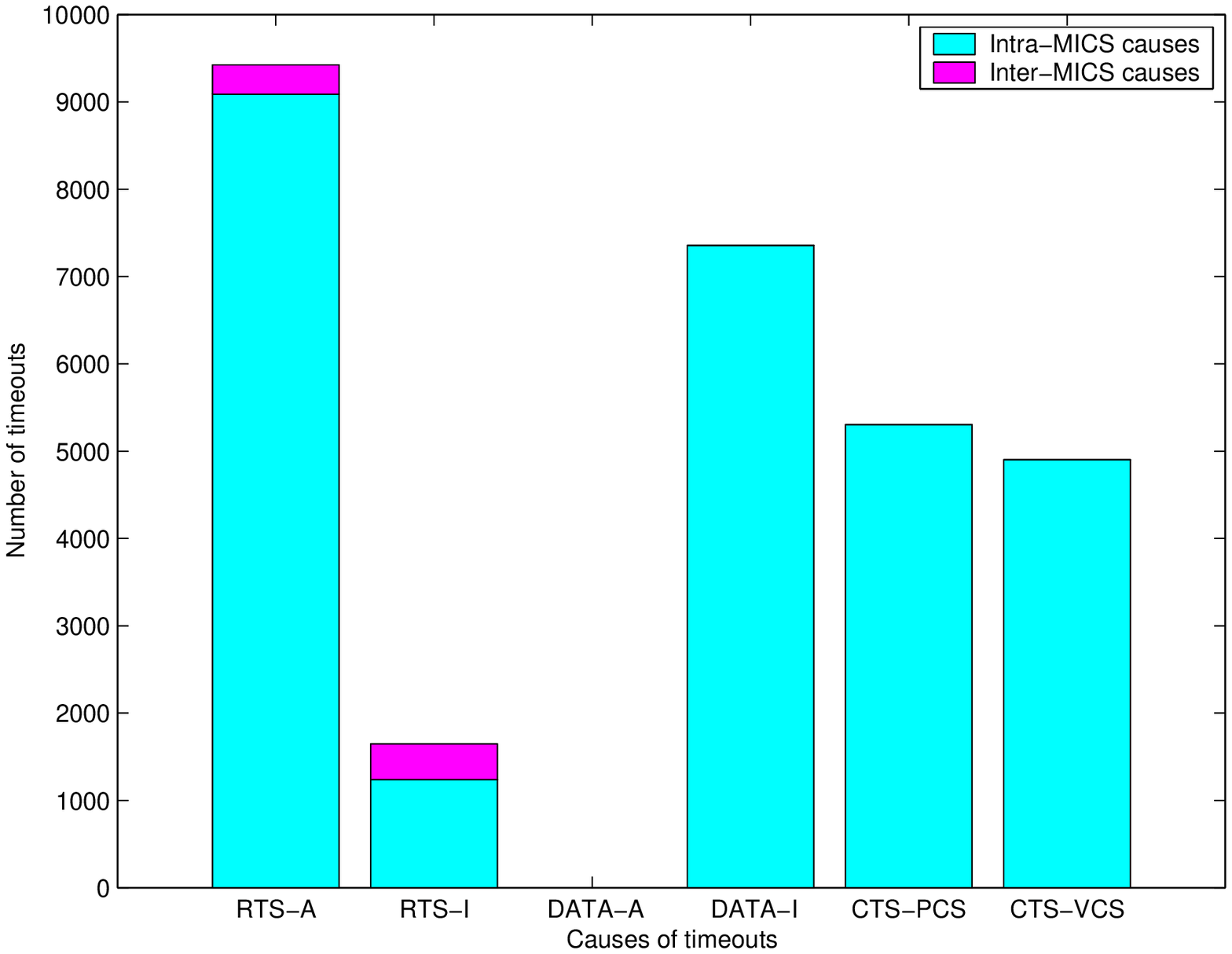}} \quad
        }
        \mbox{
        \subfigure[RTS Timeouts\label{fig:rtsDrop}]{\includegraphics[scale=0.5]{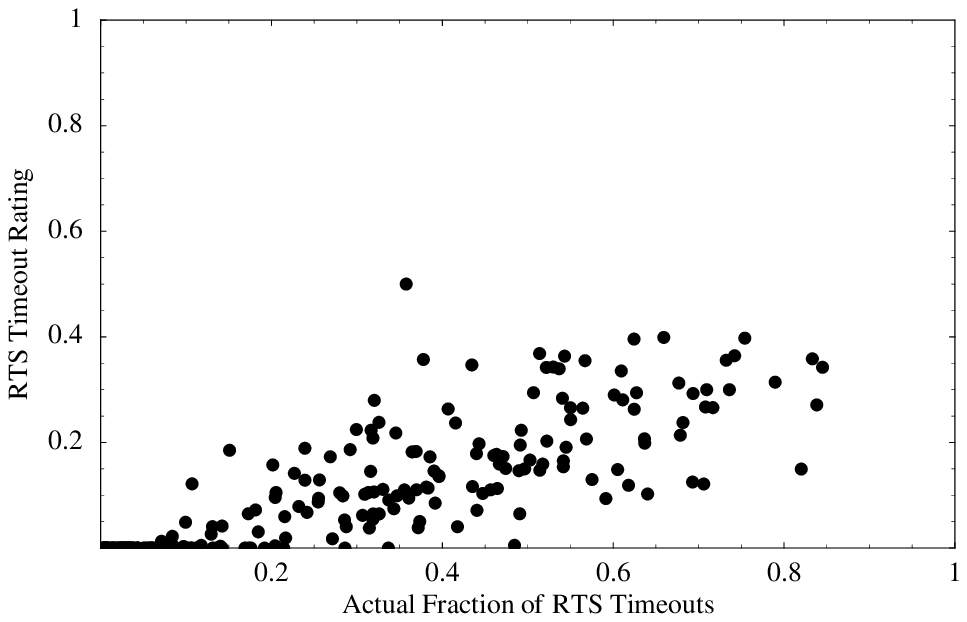}} \quad
        }
        \mbox{
        \subfigure[ACK Timeouts\label{fig:dataDrop}]{\includegraphics[scale=0.5]{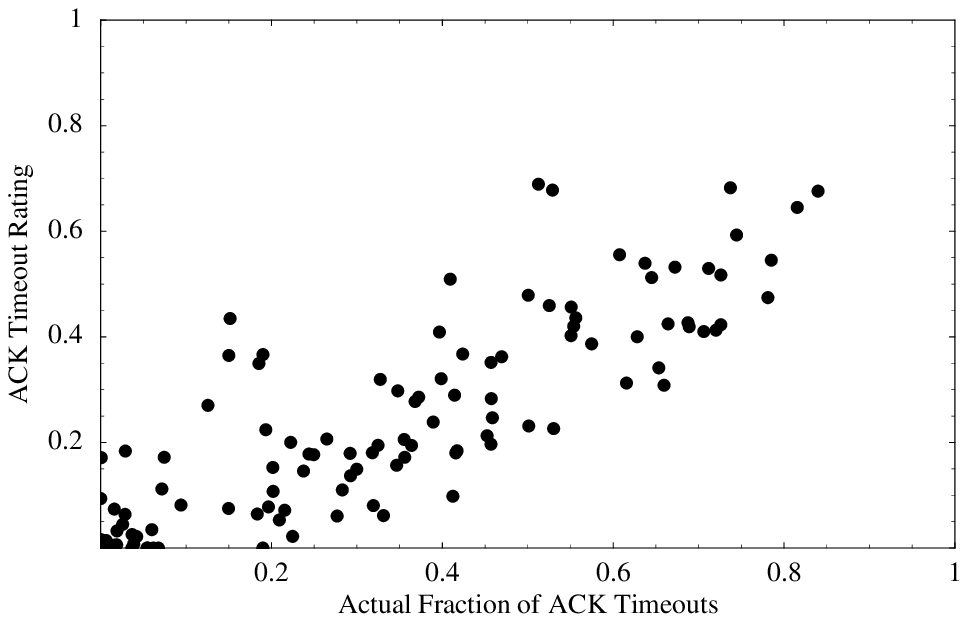}} \quad
        }
\end{center}
\caption{Validation and effectiveness of IBLR metrics}
\label{fig:schedRes}
\end{figure*}
 \begin{table*}
 \begin{minipage}[b]{0.5\linewidth}
 \centering
         \begin{tabular}{|c|c|c|c|c|}
         \hline
         \backslashbox{}{}& $\rtsRating$ & $\cal{I}$ & $\cal{T}$ & $\cal{S}$ \\
         \hline
         $R$ & 0.819 & -0.121 & 0.655 & -0.059\\
         $\rho$ & 0.872 & 0.183 & 0.529 & -0.404\\
         \hline
         \end{tabular}
 \caption{Correlation of RTS Timeouts}\label{tab:rtsCorrelation}
 \end{minipage}
 \begin{minipage}[b]{0.5\linewidth}
 \centering
         \begin{tabular}{|c|c|c|c|c|}
         \hline
         \backslashbox{}{}& $\dataRatingImp$ & $\cal{I}$ & $\cal{T}$ & $\cal{S}$\\
         \hline
         $R$ & 0.916 & -0.071 & 0.262 & -0.061\\
         $\rho$ & 0.899 & 0.110 & 0.320 & -0.558\\
         \hline
         \end{tabular}
 \caption{Correlation of ACK Timeouts}\label{tab:dataCorrelation}
 \end{minipage}
 \end{table*}

In Figure \ref{fig:rtsDrop} and \ref{fig:dataDrop}, the IBLR 
\textit{RTS Timeout Rating}($\rtsRating$) and 
\textit{ACK Timeout Rating} ($\dataRatingImp$) of the formulation 
is plotted against the fraction of RTS Timeouts 
(ratio of RTS Timeouts to the total RTS transmissions) and ACK timeouts
respectively.
% In Figure \ref{fig:rtsDrop}, the IBLR \textit{RTS Timeout Rating}
% ($\rtsRating$) of the formulation is plotted against the fraction of
% RTS Timeouts (ratio of RTS Timeouts to the total RTS transmissions)
% for all the active links observed in the simulator and similarly,
% Figure \ref{fig:dataDrop} shows the scatter plot of the IBLR
% \textit{ACK Timeout Rating} ($\dataRatingImp$) versus the percentage
% of ACK timeouts. 
Table \ref{tab:rtsCorrelation} compares the
Correlation co-efficient ($R$) and Spearman's Rank Coefficient
($\rho$) measured between the percentage of RTS Timeouts observed in the
simulator to four commonly used metrics: (1) RTS Timeout
Rating ($\rtsRating$), (2) 
Interference Level ($\cal{I}$) measurement at receiver , (3) the Busy
Time($\cal{T}$) at the receiver, and (4) Signal-to-Interference Noise
Ratio ($\cal{S}$).  Interference Level($\cal{I}$) was computed by
measuring the interference at the destination of each link by all the
active sources. The Busy Time ($\cal{T}$) is calculated as the amount
of time the destination of a link was busy due to interfering traffic
 (measured in simulation). 
The ratio of the signal strength to the cumulative interference by 
all the other sources was used to calculate SINR ($\cal{S}$).

$R$ is a statistical technique to show the strength of relationship between a pair of variables. 
$\rho$ is used to measure the ranking order strength (especially, helpful in the absence of a 
linear relationship). A $R$ or $\rho$ can take any real values between $-1$ and $1$. A value of $1$ 
indicates a perfect correlation and a value of $0$ indicates independence 
of the two values. Negative values indicate inverse relationship. 
Table \ref{tab:dataCorrelation} shows the correlation
metrics comparing ACK Timeout Rating($\dataRatingImp$), $\cal{I}$,
$\cal{T}$ and $\cal{S}$ to the percentage of ACK Timeouts obtained in
the simulator. As seen in the Figure \ref{fig:schedRes} and Tables \ref{tab:rtsCorrelation} and
\ref{tab:dataCorrelation}, there is a strong correlation between the
IBLR ratings and the simulation experiments, while the other metrics
correlate poorly. It is evident that measurement of the interference level
metrics is insufficient and the interactions of the contention
protocol has to be explicitly taken care for a strong assessment of
the channel detrimental periods.  The strong correlation of results
with IBLR metrics measured across multiple scenarios demonstrates the
independence of the metrics to various routing configurations, which
is essential (explained later in Section \ref{sec:routing}) for
comparing different routing configurations.

\section{Scheduling-aware Routing Formulation}
\label{sec:routing}
% Discussion: Analysis in chains
% In this section, we demonstrate the application of IBLR metrics
% discussed in Section \ref{sec:schedFormulation} to a globally
% co-ordinated routing scheme, which we refer as \textit{SAR} scheme in 
% the rest of the paper. Such a routing scheme serves as a example 
% to illustrate the effect of scheduling interactions in MHWNs.
In this section, we present a proof-of-concept formulation of using
the IBLR metric to improve globally coordinated scheduling-aware routing,
which we refer as \textit{SAR} scheme in the rest of the paper.
Such a routing scheme serves as a example to illustrate the effect of 
scheduling interactions in MHWNs. Accounting
for the scheduling interactions between the set of all nodes in the
network is computationally infeasible, even for a moderate sized
network.  Accordingly, a basic routing formulation is introduced that
compares the interactions between the the links which participate in
routing.  A branch-and-bound technique is then applied to the
resulting routing configuration to mutually exclude the conflicting
links by introducing additional constraints. The routing configuration
with minimal scheduling conflicts is then chosen as the
\textit{Optimal Routing Configuration}(ORC).

\noindent
\textbf{Basic formulation:}
%\label{sec:routingBasicFor}
\rednote{Vinay, should we just say we are using an MCF formulation
  that is based on or extends the existing formulations?  I am not
  sure if the level of detail here will just open up this formulation
  for questioning.  Not sure probably better to leave it.  
At first, it was not clear to me why you use
  Minimize S as the objective function, which basically just says keep
  the shortest routes.  Is that what you meant by its computationally
  infeasible to account for interference?  Please check the text below
  to make sure it is still correct.}
Routing in MHWNs for multiple connections can be formulated as a
\textit{Multi-Commodity Flow(MCF)} problem. We model the static MHWN
as a graph $G(V,E)$; each node is a vertex $v \in V$.
An edge between two vertices $a$ and $b$ represents that there can be
packet transmission from $a$ to $b$. The edge presence between $a$ and
$b$ can be inferred by the signal strength observed at $b$ when source
$a$ is transmitting. We use a power-law based model to capture the
signal strength; however, experimental means can be adopted to decide
on the edge presence. Let $(s_n,d_n,r_n)$ denote source, destination
and the rate of the $n^{\rm th}$ connection.  The rate of connection,
$r_n$, is the number of bits to be sent per unit time.  Let $C$ be the
set of connections.  Let $x_{ij}^n$ denote the flow at edge $(i,j)$ 
for the $n^{\rm th}$ connection. 

The demand for a given node $n$ for a connection
$c$ (represented as $b_n^c$) is the difference between the total
outflow from the node and total amount of inflow to the node. For all
but sources and destinations, there is zero demand.

To capture the flow at each edge, we break the flows into a set of $n$
disjoint flows, one for each connection. Let $x_{ij}^n$ denote the
flow at edge $(i,j)$ for the $n^{\rm th}$ connection. Equation
\ref{eqn:bounds} describes the limiting bound of each flow to be the
maximum rate of the connection. For a given connection, each edge can
carry a maximum load corresponding to the rate of the given
connection. The flow constraint in Equation \ref{eqn:flow} specifies
the demand requirement to be met at each node as the difference
between the outflow and inflow.
\begin{eqnarray}
%%\nonumber \textit{Feasibility Constraints:} \\
\label{eqn:bounds} 0 \le x_{ij}^{n}\le r_{n} \forall n \in C, \forall (i,j) \in E
\end{eqnarray}
\begin{eqnarray}
\nonumber b_{i}^n = \bigg(\sum_{(i,j) \in E} x_{ij}^n \bigg) - \bigg(\sum_{(j,i) \in E} x_{ji}^n \bigg)\\
\label{eqn:flow} \forall n \in C, \forall i \in N
\end{eqnarray}
A single flow per connection is used by majority of the protocols
where each edge can either carry the full traffic for a given
connection or none of it; an additional constraint is added to enforce
such an integer flow as given in Equation \ref{eqn:integerFlow}. 
This constraint can be relaxed in multi-path~\cite{ref:das_01_multipath} 
formulation where flow-splitting is
allowed. 
The variable $y_{ij}^n$ is a boolean variable which is set to
$1$ if the edge carries the traffic for the $n{\rm th}$ connection and
$0$ otherwise.
\begin{eqnarray}
\label{eqn:integerFlow} x_{ij}^n = r_n \cdot y_{ij}^n &~~~~\forall n \in C, \forall (i,j) \in E
\end{eqnarray}
Equations \ref{eqn:bounds}, \ref{eqn:flow} and 
\ref{eqn:integerFlow} form the basic
feasibility constraints of classical MCF formulation. These 
constraints do not restrict the flows
to smaller number of hops when one is available. 

Let
 \textit{Signal}($S_i$) denote the sum of all incoming and outgoing flows carried by the
 node as per Equation
 \ref{eqn:signal}. Minimizing the total signal carried by all 
 nodes reduces the number of active nodes, thus limiting choices to short 
 paths. This can be achieved by an objective function that 
\textit{minimizes} the sum of signals across all nodes 
($\sum_{i \in V} S_i$). 
% An objective function \ref{eqn:objective} is introduced for this purpose.
\begin{eqnarray}
\label{eqn:signal} S_i = \displaystyle \sum_{n \in C} \Bigg(\sum_{(i,j) \in E} x_{ij}^n + \sum_{(j,i) \in E} x_{ji}^n \Bigg)&\forall i \in N
\end{eqnarray}
% \begin{eqnarray}
% \label{eqn:objective} \text{Minimize} \displaystyle \sum_{i \in V} S_i
% \end{eqnarray}
\bluenote{Vinay: The objective restricts to shorter hops and not the 
shortest hops, thus avoiding constrained shortest path problem. So I
have altered the next sentences}
Using the simple objective function above, we are making a computational
efficiency tradeoff for illustrative purposes.  Linear programming
solution time for a such a simple objective is much faster than a more
useful metric that considers, for example, minimizing interference. 
%We believe that the quality of
%the solution will improve with such an advanced objective function.  
%However, the solution time for
%this formulation will be high, making its use as the basis of a
%branch-and-bound algorithm expensive.  
We intend to explore such
objective functions in the future.  We do use the MCF formulation with
minimizing interference as the objective function (but without scheduling effects) as one of the baseline comparisons to demonstrate the effect of accounting for scheduling.
\\\\
\noindent
\textbf{Accounting the scheduling interactions:}
%\label{sec:routingSchedEffects}
The active links obtained from the basic formulation are ranked
according the IBLR metrics (\textit{RTS Timeout} and \textit{ACK
  Timeout}) as discussed in Section \ref{sec:schedFormulation}.
Conflicting links are mutually excluded by adding constraints to the
formulation and the obtained routing configuration is re-examined.
%Accessing the overall quality of the routing configuration, constraint addition criteria and convergence decisions are discussed below.

The overall quality of this routing configuration is combined into a
single quality metric, \textit{Configuration Interaction Metric(CIM)},
which can be compared against other routing configurations. In this
study, we obtain the CIM by taking the mean of the RTS and ACK timeout
ratings of all links. Interactions to be addressed while choosing a 
suitable CIM under multiple connection scenario is one of the focus 
of our future work.

The conflicting links for a given link $(s,d)$ that lead to the RTS
timeouts can be identified by the union of the \textit{Unsafe Links}
(${\edgeUnsafe}^{C}_{(s,d)}$) across all the MICS. The links that can
lead to an ACK timeout for the link $(s,d)$ can similarly be
identified by $\RTSCorruptImp$, $\CTSCorruptImp$ and $\ACKCorrupt$.
The link with the highest rating (which has a high vulnerability for
RTS/ACK Timeouts) is identified in the given routing configuration. A
constraint (Equation \ref{eqn:conflictConstraint}) to the link $(s,d)$
and its conflicting links $(s_1,d_1)$ is added for mutual exclusion of
the links and the MCF is re-evaluated.
\begin{eqnarray}
\label{eqn:conflictConstraint} x_{sd}^c + x_{s_1d_1}^c \le r_n
\end{eqnarray}
The branching stops when either all the links have the RTS/ACK
rating lesser than a given threshold or when no more shortest path
routes(obtained for a single time from Breadth First Search algorithm) 
are available. The best CIM and its route configuration is chosen as 
the ORC.  With more flexible objective functions that allow path 
stretch, we believe that even better solutions can be found. This is 
a topic of future research.

%\\\\
%\noindent
%\textbf{Multi-hop Extensions for SAR:}
The IBLR metric assumes that each active link has traffic to transmit.
Under low traffic scenarios or due to multi-hop self-contention
effects (where downstream hops are dependent on upstream hops for
their data), this may not be the case.  The low-traffic
case is naturally accounted for in the MHWN model.  % While, this is true in an ideal multi-hop scenario where the
% pipelining of packets is well-balanced, many chains suffer from high
% traffic at the initial hops than the later hops. Initial hops of a
% connection are less likely to be conflicted by later hops. This
% imbalance of conflicts in a multi-hop chains would require accounting
% for amount of traffic at the conflicting node. We conjecture that
% under lesser number of connections, this effect would be pronounced
% because of the smaller number of high-traffic nodes contending for the
% channel.
However, extensions to model the self-interference effects in chains
are needed to better estimate link quality in such scenarios; this is
an area of future work.
% The choice of the CIM should be generic across all the routing
% configurations to aid comparison. Combination mechanisms include (1)
% minimizing the maximum rating(RTS or ACK Timeout), (2) weighing the
% more critical ACK timeouts more than the RTS, and (3) The estimated
% traffic at the link based on its hop-position in the
% connection(initial hops have higher traffic than later hops). The
% choice of CIM can affect the choice of ORC and is our future
% research direction.  
% Another extension of the model  for
% connections with different data packets sizes and unequal sending
% rates. We believe that the formulating these behaviors will be helpful
% for analysis of TCP traffic. % The model can also be extended to a joint
% interference-and-scheduling aware formulation would increase the
% number of parallel transmissions while reducing the conflicts among
% the links under contention based protocols. Addressing dependencies in
% such formulation would provide a fundamental insight on the behavior
% of MHWNs.

% \input{branchcut.tex}
\begin{figure*}[ht]
\begin{center}
        \mbox{
        \subfigure[Throughput Study\label{fig:through.grid}]{\includegraphics[scale=0.30]{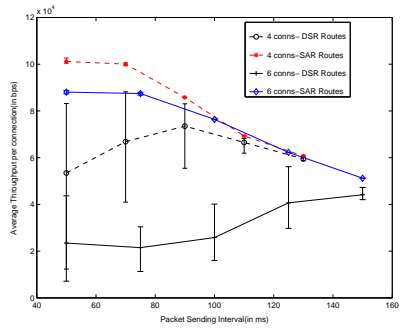}} \quad
        }
        \mbox{
        \subfigure[Connection metrics\label{fig:perf.grid}]{\includegraphics[scale=0.30]{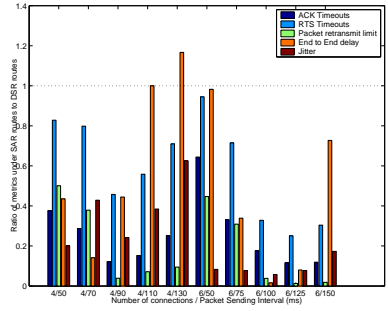}} \quad
        }
        \mbox{
        \subfigure[Performance in Random Topology\label{fig:rand.metrics}]{\includegraphics[scale=0.30]{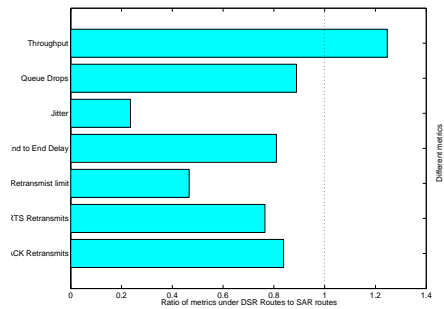}} \quad
        }
        \mbox{
        \subfigure[Comparing to Interference Aware Routing\label{fig:iferCompare}]{\includegraphics[scale=0.30]{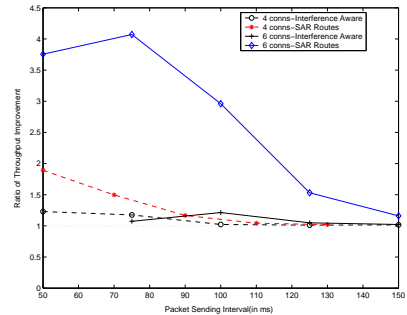}} \quad
        }
\end{center}
\caption{Performance of Scheduling-aware Routing(SAR)}
\label{fig:res.grid}
\end{figure*}

\noindent
{\bf Performance Evaluation of SAR}: In this section, we evaluate the
performance of SAR with standard routing schemes. A simple grid
topology is first studied to illustrate the effect of the scheduling.
The DSR routing protocol was used to compare the scheduling
effectiveness.  Since SAR uses static routes, and to provide a fair
comparison, eliminating routing overhead and false disconnections, we
allowed DSR to use static routes (selecting the most commonly used
shortest path found by DSR).

A 4 and 6 connection scenario was analyzed in a 6$\times$6 grid
topology. The sending rate was altered to observe the effects at
various traffic loads. Figure \ref{fig:through.grid} shows the
throughput analysis. The error-bars indicate the variation of the
throughput across different seeds. A performance improvement of around
4x and 2x was observed with 4 and 6 connections respectively.  Note
that under low sending rate (higher sending intervals), the scheduling 
effects play a less
important role. Under high traffic loads, it can be observed that even
the \textit{maximum} throughput of the DSR routing scheme is
significantly lesser than a scheduling aware routing scheme.  
% There is
% also a very large variation in throughput because DSR only uses hop
% count and not link quality in selecting routes).
% When
% compared with DSR protocol (without static routes), the throughput of
% SAR was on an average 2x and 8x greater for 4 and 6 connection
% scenario, respectively. All the individual throughputs of SAR were
% greater than the ones observed in DSR protocol (comparable at larger
% sending intervals).
Figure \ref{fig:perf.grid} shows the ratio of various connection
metrics for SAR compared to the idealized DSR routes. 
A pronounced improvement was found in the majority of the metrics, 
especially the collision metrics like RTS Timeout, 
ACK Timeout and the packet drops due to retransmission limit. 
The ratio of standard routes' connection
metrics to the SAR metrics under a scenario of 8 random multi-hop
connections in a random deployment of 144 nodes in a
1600m$\times$1600m area is shown in Figure \ref{fig:rand.metrics}.
\\
\noindent
\textbf{Comparison with Interference Aware Routing:}
\label{sec:routingifer}
The effectiveness of the scheduling interactions was compared with
\textit{Interference Aware Routing} (IAR) schemes. Existing IAR schemes 
follow the \textit{Protocol Model}
~\cite{ref:jain_03_interference}. % Comparing a
% SINR model to Protocol Model results in unfair comparison. Hence, 
% the improvements observed with the standard routes under Protocol Model
% and SINR model were compared with IAR and SAR scheme. 
Figure \ref{fig:iferCompare} compares the ratio of the throughput of IAR and 
SAR to the counterpart standard routes under different sending rates. 
It was observed that under a reasonable traffic, IAR routes do not
account for scheduling, and therefore perform significantly worse than
SAR routes.
\section{Concluding Remarks}\label{sec:conclude}

In this paper, we first demonstrated that aggregate metrics of
link capacity such as observed channel busy time  
are only effective in estimating the upper bound of link
performance.  Examination of low-level scheduling effects demonstrates
that MAC interactions play a critical role in how interference is
manifested.  The key insight is that destructive behavior at the MAC
level arises from sources that interfere but are not prevented from
transmission by the MAC protocol.  Thus, it is desirable to 
avoid links that experience mutually destructive interaction.

We used this observation to develop a methodology for ranking active
links based on their interactions with other active links.
Specifically, we identify the sets of sources that can transmit
concurrently per the rules of the MAC protocol.  With the exception of
rare race conditions, collisions occur due to interactions within such
sets.  Based on an analysis of destructively interacting sources in
independent sets, we develop an estimate for the expected RTS
timeouts and expected ACK timeouts.  We show that these estimates,
which we call Interference Based Link Rating (IBLR) correlate strongly
with observed packet drops.  
We demonstrate the effectiveness of IBLR rating by using them in a
linear programming formulation of traffic engineering in static MHWNs.
We show that capturing the scheduling effects leads to considerable
improvement in performance of the derived routes, even for the small
size scenarios we considered. If these results can transition to
realistic environments, they have important implications for mesh
network traffic engineering and provisioning.  Further, identifying
the nature of collisions that influence the quality of the links
provides a starting point for distributed algorithms that capture
these interactions in a distributed environment.

% Our future work includes continued refinement of the model, which uses
% coarse approximations in several points. More interestingly, we seek
% to apply these lessons to realistic protocols. We believe that the
% concept of an MICS and unmanaged collisions within them applies in
% realistic environments. However, due to the changing nature of the
% wireless channel, the formulation may need to be expressed
% probabilistically. In addition, due to the extremely encouraging
% initial results, we seek to apply these techniques to other instances
% of MHWNs like mobility support. Thus, distributed
% formulations of our solutions need to be developed. The fact that
% interfering MICS members are geographically constrained provides a
% starting point for developing such protocols.

\bibliography{references}
\bibliographystyle{IEEEtran}
\end{document}